\documentclass[conference]{IEEEtran}
\usepackage{url}
\IEEEoverridecommandlockouts

\usepackage{cite}
\usepackage{amsmath,amssymb,amsfonts}
\usepackage{algorithmic}
\usepackage{graphicx}
\usepackage[export]{adjustbox}
\usepackage{textcomp}
\usepackage{comment}
\usepackage{xcolor}
\usepackage{multirow}
\usepackage{stfloats}
\graphicspath{{../figs/}}
\usepackage{caption}
\usepackage{listings}
\usepackage{subcaption}
\DeclareGraphicsExtensions{.png}
\usepackage{float}
\def\BibTeX{{\rm B\kern-.05em{\sc i\kern-.025em b}\kern-.08em
    T\kern-.1667em\lower.7ex\hbox{E}\kern-.125emX}}

\newcommand{\updates}[1]{\textcolor{black}{#1}}

\begin{document}

\title{Autopsy of Ethereum's Post-Merge Reward System}

\author{
    \IEEEauthorblockN{Mikel Cortes-Goicoechea}
    \IEEEauthorblockA{
    \small{Barcelona Supercomputing Center}\\
    Barcelona, Spain \\
    mikel.cortes@bsc.es}
    \and
    \IEEEauthorblockN{Tarun Mohandas-Daryanani}
    \IEEEauthorblockA{
    \small{Barcelona Supercomputing Center}\\
    Barcelona, Spain \\
    tarun.mohandas@bsc.es}
    \and
    \IEEEauthorblockN{Jose Luis Muñoz-Tapia}
    \IEEEauthorblockA{
    \small{U. Politécnica de Catalunya}\\
    Barcelona, Spain \\
    jose.luis.munoz@upc.edu}
    \and
    \IEEEauthorblockN{Leonardo Bautista-Gomez}
    \IEEEauthorblockA{
    \small{Status.im}\\
    Barcelona, Spain \\
    leo@status.im}
}

\maketitle

\begin{abstract}
Like most modern blockchain networks, Ethereum has relied on economic incentives to promote honest participation in the chain's consensus. The distributed character of the platform, together with the "randomness" or "luck" factor that both proof of work (PoW) and proof of stake (PoS) provide when electing the next block proposer, pushed the industry to model and improve the reward system of the system. With several improvements to predict PoW block proposal rewards and to maximize the extractable rewards of the same ones, the ultimate Ethereum's transition to PoS applied in the Paris Hard-Fork, more generally known as "The Merge", has meant a significant modification on the reward system in the platform. In this paper, we aim to break down both theoretically and empirically the new reward system in this post-merge era. We present a highly detailed description of the different rewards and their share among validators' rewards. Ultimately, we offer a study that uses the presented reward model to analyze the performance of the network during this transition.    
\end{abstract}


\section{Introduction}
\label{sec:introduction}

Ethereum~\cite{eth-whitepaper} has soundly consolidated in the Web3 and blockchain ecosystem for its widely decentralized network with more than $5.000$ active nodes~\cite{cortes2021discovering}, and its multipurpose Ethereum Virtual Machine (EVM)~\cite{hildenbrandt2018kevm}. Commonly popularized for being the pioneer blockchain able to process \emph{Smart Contracts}~\cite{wohrer2018smart}, it currently handles above $1$ million transactions per day from $600$ thousand active accounts~\cite{oliveira2022analysis}. As with any other blockchain in the Web3 space, to support this decentralized platform, Ethereum relies on an incentive program to reward honest behaviors between the participants. 

Starting its journey as an operative network in 2015, Ethereum chose Proof of Work (PoW) as its consensus mechanism. In this PoW-powered platform, participating nodes (commonly known as miners) competed with each other to become the next block proposer. Nodes in the network built block candidates by adding user transactions from the public \emph{Mem-pool}. Awarded with a generous tip in Ethereum's token Ether (ETH) for the first one publishing a valid block, miners would parallelly iterate over multiple nonce values\footnote{Nonce values in PoW-based blockchains is a field in the head of the block structure that miners can indistinctly modify to satisfy the hash target requirements of the chain.} until the hash value of the entire block satisfies the target required by the network.

Initially, as part of the incentives that the block proposal had until EIP1559~\cite{roughgarden2020transaction} was applied, block proposers were also granted with the fees coming from processing the transactions included in the block plus a dynamically adjusting base reward. 
However, as long as the network has proved to be resilient, powering the consensus through PoW required huge and unnecessary power consumption. As one of the biggest upgrades in the platform's lifetime, Ethereum embraced a long journey to shift its consensus mechanism from PoW to a more sustainable Proof of Stake (PoS) \cite{saleh2021blockchain}. The transition started with the launch of a side chain with the single purpose of achieving consensus. To participate in the Beacon Chain or Consensus Layer (CL), users now have to deposit in a publicly available smart contract 32 ETH to activate a validator. The beacon chain decides who is the next block proposer (among many other things) using a randomness generator algorithm called RANDAO~\cite{randao-code}. The beacon chain is organized in epochs, where each epoch contains 32 slots of 12 seconds, making a whole epoch last 6.4 minutes. Each slot will serve as a time range where the RNG algorithm decides which validator can propose the block. Removing the competition between nodes to become block proposers reduced by 99.95\% the total power consumption of the network, with the significant drawback of adding extra complexity to reaching consensus. 

The beacon chain stores all the information regarding the status of the validators, from when they are activated, to their interaction with the network when performing their duties, including their balance. To reach a consensus and maintain the finalization of the chain, these validators now have to perform a set of tasks at every epoch of the chain. Also known as duties, validators are grouped by committees, which have the major task of voting on the correctness of the proposed block (if proposed) from the slot they are assigned to. These votes are also known as attestations and are included in the beacon blocks. Validators are also grouped by sync committee, a small set of 512 validators that help sign headers of blocks to prove their validity to light clients that won't need to sync the entire chain. 
Based on the correctness of all these duties, validators get rewarded as an incentive to well-behave in the consensus, making honest actions in the network more economically lucrative than dishonest ones. 

Both parallel chains had co-existed since December 2020, when the beacon chain was launched, until September 2022, when the chain in charge of processing transactions became consensus-dependent from the beacon chain. In this event called ``The Merge"\cite{paris-hardfork}, active validators in the consensus layer will also be the ones proposing blocks in the EVM chain (from now on called Execution Layer (EL)). With the successfully accomplished transition from PoW to PoS, there are some significant changes in the new reward mechanism of the multipurpose platform. Since validators now propose both blocks, they get the CL and the EL rewards. However, from the EL rewards, validators only earn the part coming from the transaction fees, which means that there is no base reward anymore. The fact that block proposers are chosen beforehand and the mining process has been removed increases the time to include transactions in a block. Which ultimately increases the chances of a proposer gaining more rewards
by filtering and ordering the transactions. 


This paper presents a previously unseen approach to measure the transition from PoW to PoS in a live-working blockchain network. Our approach uses both empirical and theoretical methods to compare and score how well validators accomplish their duties. Moreover, we present a comparison in performance based on rewards for entities that represent a large set of validators in the network.

The paper is organized as follows: Section \ref{sec:state-of-art} introduces the state of the art of the paper, going through previously done work on the topic, Section \ref{sec:methodology} introduces all the rewards-related terms, theoretical methods, and the methodology that we followed, Section \ref{sec:discussion} presents the discussion over the results we gathered from the study, and Section \ref{sec:conclusion} summarizes all the highlights of the study. 

\section{State of the Art}
\label{sec:state-of-art}


Blockchain-based networks adopted the peer-to-peer philosophy to consolidate platforms able to store a public ledger. Generally, these public ledgers use incentive systems to promote honest participation in the consensus. This way, for most users, honest participation in the network is more rewarding than misbehaving. Ethereum also uses the same system to promote the chain's finalization (reach a consensus on which chain to follow).

The unknown nature of block rewards in Ethereum's execution layer has attracted many experiments that tried to modelize it. Authors in \cite{salimitari2017profit} already introduced the apparition of mining pools for the Bitcoin \cite{nakamoto2008bitcoin} network. Pools where individual miners unified efforts to split rewards and costs, making them more stable over time. Other works like \cite{simon2021block} presented an approach that, using linear and polynomial regressions, could predict the mining reward of the next Ethereum block to some degree. More focused on protocol optimizations, other works like \cite{albert2020gasol} introduced an intensive study highlighting the under-optimized storage patterns that most smart contracts use. Furthermore, they propose a tool that helps optimize gas usage when developing smart contracts.  

Showing a clear interest in predicting and maximizing the rewards extracted by each block, authors in \cite{daian2019flash} introduce the lack of fairness that decentralized exchanges and arbitrage bots introduced to the platform by using the usage of sandwich attacks \cite{qin2022quantifying} to front run users transactions. The authors present a novel idea of block-building audition or bid, where they introduce a set of techniques that optimize the reward extracted from each built block (also known as Miner-Extractable Value (MEV)), i.e., transaction ordering.  

The launch of the consensus layer and the shift of Ethereum to PoS, resulted in a way more sustainable network with over 400k validators distributed between 5k nodes in 90 different countries \cite{cortes2021discovering} running mostly five different software that adjusts to all kinds of users' need while increasing the overall resilience of the network ~\cite{cortes2021resource} to software bugs. However, despite the community's efforts to endorse a fully decentralized blockchain platform, authors in \cite{cortes2021discovering} identify a certain degree of centralization in terms of geographical location and used software. 

This paper aims to complete the literature surrounding Ethereum's rewards from the consensus layer perspective. In our study, we present a novel method to evaluate the performance of validators by \updates{comparing their on-chain rewards with the maximum theoretical} one they could obtain, identifying which are the most rewarding duties of the platform. We introduce the results of applying this method to Ethereum's transition from PoW to PoS during the merge. Furthermore, we present the decentralization degree that the current Ethereum network has by analyzing the distribution of validators among staking entities in the network.

\section{Methodology}
\label{sec:methodology}

\begin{figure}
    \centering
    \includegraphics[width=\linewidth]{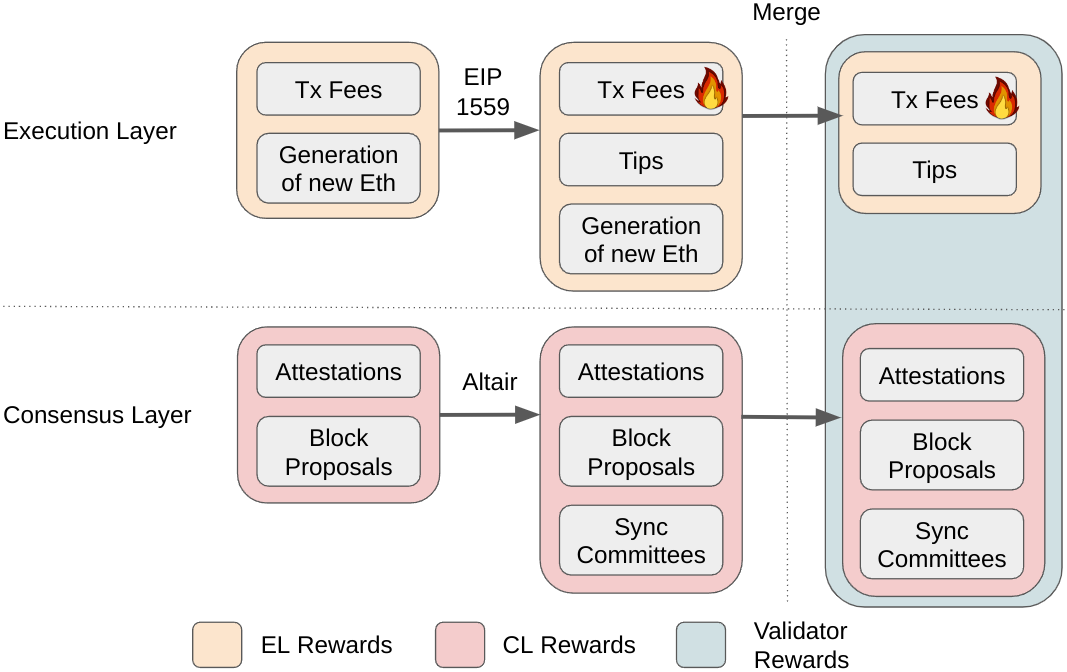}
    \caption{\updates{Evolution of rewards for different layers in Ethereum.}}
    \label{fig:rewards-evolution}
\end{figure}

Economical remuneration is a highly used incentive to recompense honest behaviors in blockchain scenarios. However, building a sustainable economic rewards system with a solid game theory \cite{liu2019survey} requires considerable effort. This section explains the different reward types that Ethereum miners (previously) and Ethereum validators can earn from their honest interaction with the network, \updates{describing in detail the evolution of rewards displayed in Figure \ref{fig:rewards-evolution}.}

\begin{figure}
    \centering
    \includegraphics[width=\linewidth]{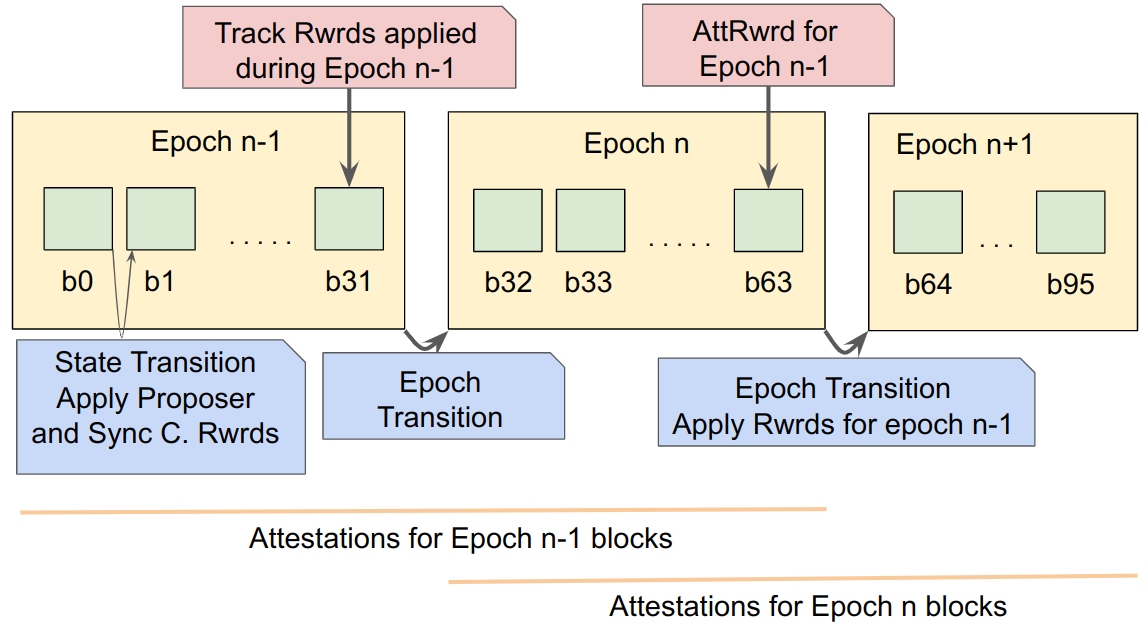}
    \caption{Rewards extracting schema followed by the state analyzer tool for Epoch $n-1$ (in red: actions of our indexer tool; in blue, actions at the protocol).}
    \label{fig:epoch-processing-schema}
\end{figure}

\subsection{Execution Layer's Block Proposal Rewards}
\label{subsec:el-rewards}

As a method of preserving a fair usage of the EVM, interacting with the Ethereum EL requires paying for the needed resources. Following its close analogy to vehicles needing gas to move, users pay Gas for their interaction with the EVM to the network processing it. Thus, each interaction with the EVM requires paying Gas for block space for the user's transactions. With the introduction of EIP 1559~\cite{roughgarden2020transaction} in the \emph{London Hard Fork}\cite{london-hardfork} \updates{(see Figure\ref{fig:rewards-evolution})}, the fees that a user pays to include a transaction in a block are divided into three different values:
\begin{enumerate} 
    \item Gas Limit: Upper limit of Gas that the entire operation may consume, protection against malicious or broken smart contracts.
    \item Base Fee per Gas: Minimum price per gas unit to allocate a transaction in a block. Base Fee dynamically adjusts based on network congestion; the more demand for block space, the higher the Base Fee.
    \item Priority Fee per Gas: colloquially known as the tip users pay to prioritize their transactions.
\end{enumerate}
Leaving the total transaction as the product between the \emph{GasLimit} and \emph{BaseFee+Tip}.

In this new Gas system, block proposers in the execution layer no longer receive 100\% of the Gas paid by the user. Currently, the aggregation of base fees included in each block is burned. As a result, since EL block proposals do not create new ETH in a post-merge scenario, EL block proposers only get the aggregation of the \emph{Gas Tips} included in the block.

\subsection{Characterization of Attestation Rewards}
\label{subsec:attestation-rewards}

As part of each validator's duties in the network, once every epoch, all active validators are asked to participate in the consensus system by voting on whether a specific block is correct or not (understanding missing blocks as not correct ones). With this method, the network agrees and validates the data in the chain (finalization). As a result of performing this task, validators earn attestation rewards.

The whole set of validators is divided into beacon committees in each epoch of the beacon chain. However, each committee is assigned one slot per epoch. Thus, each validator is randomly assigned to attest to a single committee at a single slot, having a total of 32 slots to do it.
After each validator has shared its vote with its respective beacon committee, all votes in that committee are then aggregated into a single attestation. If some validators send their vote later, another attestation is created with these votes, so no votes are left behind if they were sent inside a window of 32 slots (one epoch) after the slot they are attesting to. Moreover, the attestation or vote that each validator produces contains three main flags that determine the ``quality" of the attestation:
\begin{itemize}
    \item Source: the hash of the justified checkpoint\footnote{Checkpoints in the CL represent the Beacon State root of the epoch's first slot, including the result of the state transition from the previous epoch.} at the moment the attested block was proposed.  
    \item Target: the hash of the first block at the epoch the attested block was proposed.
    \item Beacon block root: hash of the attested block.
\end{itemize}

After the \emph{Altair Hard Fork}\cite{altair-hardfork}, each flag has a respective weight, determining the final reward per flag. Ultimately, the summary of the product between these three flags and their weights and a \emph{BaseReward}\footnote{Specific value that depends on the validator's effective balance and total active balance at a given epoch.}, represent the attestation reward a validator will receive for its attestation to a given block.

From Ethereum's Consensus specs \cite{eth-cl-specs}, the rewards for each attestation are directly calculated using the following formulas. Please note that for the sake of clarity in the paper, we will summarize some of the aspects of the procedures to the basics (in the code implementation, for a better resolution between calculus, the formulas are calculated per units of the validator's effective balance).
\begin{equation}
    BaseRwrd=\frac{EffBalance * BaseRewardFactor}{\sqrt{TotalActBalance}}
\end{equation}
\begin{equation}
    FlagRwd=\frac{FlagWeight}{WeightDenom}*BaseRwrd*\frac{AttBalance}{TotActBalance}   
\end{equation}
\begin{equation}
    AttestationReward=\sum{FlagRwrd}
\end{equation}
Where: \emph{EffBalance} is the effective balance of the attesting validator, \emph{BaseRewardFactor} is a constant of value $64$, \emph{FlagWeight} and \emph{WeightDenom} are constants that vary between flags, \emph{AttBalance} is the sum of effective balances of all attesting validators, and \emph{TotActBalance} is the sum of the effective balance of all active validators.

\subsection{Characterization of Sync Committees Rewards}
\label{subsec:sync-rewards}

Since the \emph{Altair Hard Fork}~\cite{altair-hardfork}, the Ethereum network has introduced the idea of sync committees, a group of 512 randomly selected validators who sign new block headers every slot. Light clients can use these headers to trust which blocks have been validated without fully downloading and processing the beacon chain.
For a total of 256 epochs (8192 slots), each validator participating in a sync committee receives an extra reward: the sync committee reward. However, unlike attestation rewards, this reward is added at the state transition\footnote{The state transition corresponds to the act of including all the duties in the beacon chain in the parent Beacon State to produce the new Beacon State for the next slot. Includes balance modifications, slashing, etc.} of every slot, not at the epoch processing. After 256 epochs, the list of validators in the sync committee is refreshed.

For every slot that a validator correctly participates in the sync committee (by signing the new block headers and sharing this information with the other participants of the sync committee), its balance will be immediately updated at that slot. However, if the validator does not participate correctly, it will suffer a penalty for missing its duty.
The sync committee rewards per block can be summarized as follows:
\begin{equation}
TotSyncComRwrd=TotBaseReward*\frac{SyncRwrdWeight}{WeightDenom}
\end{equation}
\begin{equation}
    IndvSyncComRwrd=\frac{TotSyncComRwrd}{SlotsPerEpoch*SyncComSize}
\end{equation}

Where: \emph{TotSyncComRwrd} is the sync committee reward for the whole committee per epoch, \emph{TotBaseReward} is the sum of the \emph{BaseReward} of all the active validators, \emph{SyncRwrdWeight} and \emph{WeightDenom} are constants related to the sync committees, \emph{IndvSyncComRwrd} is the individual reward per validator per slot, \emph{SlotsPerEpoch} is a constant of 32 slots per epoch, and \emph{SyncComSize} a constant of 512 validators per sync committee. 

\subsection{Characterization of Block Proposals}
\label{subsec:block-proposals}

For every epoch, a total of 32 validators are randomly chosen to be block proposers. Each block proposer will have the duty of proposing a block in one of the slots of the epoch. In the opposite direction to the constant and stable attestation rewards, block proposal rewards are way more sporadic due to the randomness of the block proposer election. With a current total of 470k active validators in the network, the chances of being elected as a block proposer are around one every two months. However, the rewards linked to proposing a beacon block are high. Those rewards are coming from:
\begin{itemize}
    \item Including the attestation aggregations from validator votes that are not yet included in the beacon chain.
    \item Including sync aggregates. The block proposer is in charge of including the sync aggregate from the 512 sync committee participants.
\end{itemize}
Similarly to sync committee rewards, proposer rewards are added to the validator’s balance when the slot (and, thus, the block) is processed, being the formulas to calculate the proposer reward per each block the following:

\begin{equation}
    AttRewards=\frac{\sum{BaseReward*FlagWeight}}{\frac{(WeightDenom-PropWeight)*WeightDenom}{PropWeight}}
\end{equation}
\begin{equation}
    SyncCRwrds=\sum{\frac{IndvSyncComRwrd*PropWeight}{(WightDenom-PropWeight)}}
\end{equation}
\begin{equation}
    ProposerRewards=AttRewards+SyncCRwrds
\end{equation}
Where: \emph{BaseRewards} corresponds to every validator whose vote was included in the block, \emph{WeightDenom} and \emph{PropWeight} are constants related to the proposer rewards, and \emph{IndvSyncComRwrd} is the individual sync committee reward for each sync committee attestation included in the block.

\updates{
\subsection{Validators' Penalizations}
\label{subsec:penalization}
As an incentive to finalize epochs, Ethereum PoS validators receive rewards for their contributions when they are consistent with 66\% \cite{buterin2017casper} of the total validators (reaching consensus). However, they can also be penalized when they don’t comply with their duties or when their duties don’t match with the majority of validators. Penalties are the only protection mechanism that the platform has to defend the finality of the chain against bad actors or long time disconnections, ensuring that eventually, the network will be able to finalize again because the balance of those actors ends up below the minimum one needed to participate in the consensus.
In terms of quantifying the penalties, the PoS consensus mechanism penalizes validators when they miss-perform critical duties summarized as follows:
\begin{itemize}
    \item When a validator misses attestation flags, only the \emph{Source} and \emph{Target} flags will penalize the validator's balance. The quantity of balance that will be deducted equals the reward it would get for each flag but in negative. This ensures that one day of correct attestations means more rewards than the deducted ones for an inactivity day.
    \item Missing block proposals are bad enough for the validators as they represent a high reward in a single slot. The rest of the network is in charge of attesting for an empty slot. Thus, no extra penalty is applied to the validators that missed a block proposal.   
    \item The network doesn't have much mercy on validators when they miss a sync committee duty. Because they are essential for light clients to verify the validity of the latest epochs. The entire reward that the validator could have had, will be deducted from its balance when sync committee duty is missed. 
\end{itemize}
The previously introduced penalties refer to the duties of a validator that were, according to the consensus, wrongly done or missed. However, there are a set of actions a validator could easily do that could create conflicts inside the network. For this reason, the consensus protocol has a set of rules that, whenever any validator breaks them, it is subject to slashing. The slashing can be triggered when a validator performs two times the same duty, i.e., voting with two different attestations to the same slot, as it can divide the network in different forks depending on which attention was received first. The slashing of a validator presents the exit of the validator from the list of active ones and three concurrent balance modifications:
\begin{itemize}
    \item The penalization of the validator whose balance gets decreased:
    \begin{equation}
    SlashingPenalty=\frac{EffBalance}{32}
    \end{equation}
    \item The reward for the validator that proposed the block that includes the slashing, whose balance increases:
    \begin{equation}
    PropSlashRewrd=\frac{EffBalance*PropWeight}{512*WeightDenom}
    \end{equation}
    \item The reward of the \emph{Whistleblower} that reported the incident, whose balance increases: 
    \begin{equation}
    WhistleBReward=\frac{EffBalance}{512}-PropSlashRewrd
    \end{equation}
\end{itemize}
}
\subsection{Maximum Extractable Reward (MER)}
\label{subsec:mer}

The presented formulas in the above sections describe Ethereum's current reward system at the time that we are writing this paper (post-Paris Hard Fork~\cite{paris-hardfork}). As a reminder, for a healthy performance of the beacon chain (keep finalizing states), only 66\% of the validators need to actively participate in the consensus on new blocks. This means that even though the network achieves resilience to validator churn\footnote{Validators that go offline for a period of time and come back afterward (due to client updates, internet shortage, among others).} the network is still able to finalize states in the chain. 
We understand as participating validators, the ones that are actively attesting in their committees; however, in the last instance, all the rewards that attesting validators, sync committee members, and block proposers receive also depend on the quality of their duties (i.e., an attestation might have 1/3 flags correct and still be considered as participation to the consensus).

This said, and as part of our contribution to this paper, we identified that we could theoretically qualify and compare the max reward each of the validators could achieve at every given epoch with the one they actually achieved. From now on, we will refer to this theoretical max reward as Maximum Extractable Reward (MER).

\subsection{Beacon State Analyzer}
\label{subsec:bstate-analyzer}

As a method to index the status of each validator in the network and the data included in the beacon chain, we developed an open source tool~\cite{state-analyzer} that can index the individual duties per validator per epoch, the quality of these duties (i.e., if validators missed a block proposal, the number of flags that were successfully voted, etc.), and the max attestation and sync committee (if the validator during that epoch was inside a sync committee) reward that each validator got in each epoch.

\updates{The tool can index in a PostgreSQL database metrics for any existing slot or epoch the indicated range, and it does it by requesting the last Beacon State\footnote{The Beacon State is the main data structure that represents the beacon chain’s status at every slot, including the entire list of validators, their balances, and duties.} of each epoch in the indicated time range from a trusted synchronized Beacon Node}. In the previous sections, we already introduced that there are two ways of applying rewards to the validators: during the state transition at the end of each slot (block proposal and sync committee rewards) or during the epoch transition that defines the first state of the next epoch (the state transition in the last slot of the epoch, where attestation rewards are applied). 

Since attestations can be included 31 slots after the block they are referring to, the tool processes the rewards following the schema in Figure \ref{fig:epoch-processing-schema}. The tool chronologically analyzes the states, tracking the proposer and sync committee rewards applied during epoch $n-1$ at the last epoch slot (slot $31$ in the figure). Once we have already distinguished the rewards, we process the next state, which is the state of the last slot in epoch $n$ (slot $63$ in the figure). From the second state that we analyze until the end of the given time range, we can index attestation rewards along the sync committees and block proposals. In the last slot of epoch $n$, we can already measure which are the duties that were accomplished in relation to epoch $n-1$ and which ones were not, which let us rate how well each validator performed and which would be its max reward for that particular attestation. Although it might be a bit misleading, these attestation rewards we can already compute won't appear in the validator's balance until the epoch transition is applied.

\subsection{Validator to Staking Pools Mapping}
\label{subsec:staking-pools}
We have already introduced the steps and requirements to become a validator in Ethereum's PoS. However, collecting 32 ETH (for optimal performance) and running the required software 24-7 at home is not possible for every user. Either the lack of funds or technical knowledge to set up and maintain the validator online creates a large opportunity for delegated staking and the generation of staking pools. As it originally was for PoW with mining pools, the concept of staking pools consists of aggregating multiple users' funds until you can activate a validator. This type of service is generally driven by entities or organizations that are in charge of activating the validator, setting up the clients, and maintaining it.

In the Ethereum ecosystem, there are many entities. There are centralized organizations such as Coinbase, Kraken, or Binance, and decentralized entities like Lido Finance or Rocket Pool, which allow users to delegate their staking. In order to differentiate the validator performance of the wide set of staking pools, we mapped the whole set of validators to their respective staking pool (only for those ones linked to a staking pool) based on the transaction that activated the validator using the PoS smart contract. Since activating the validator in the Beaconchain requires interacting with smart contracts from a \emph{deposit address}, we inspected the whole list of PoS deposits, mapping (if possible) the \emph{deposit addresses} to any of the Ethereum addresses publicly identified from the entities.

\begin{figure}[hb!]
    \centering
    \includegraphics[width=1.05\linewidth]{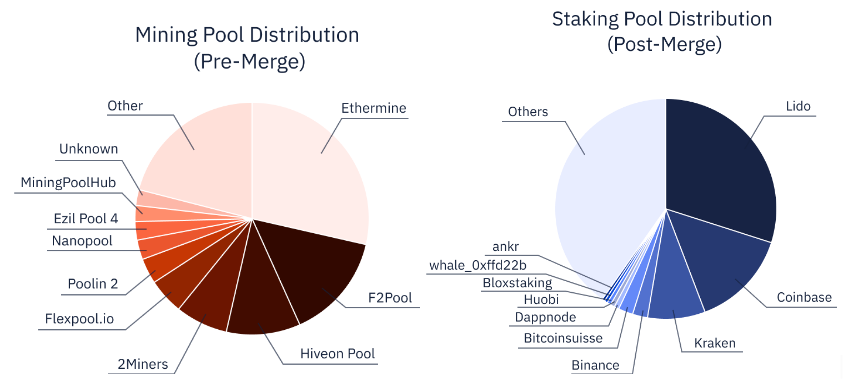}
    \caption{Block proposal's distribution per mining and staking entities for the two months of study.}
    \label{fig:staking-descentralization}
\end{figure}

\section{Discussion}
\label{sec:discussion}
In this section, we introduce the insights of the analysis we performed by indexing the rewards of all the validators active in the network between epochs $139875$ and $153875$, \updates{which are introduced in larger detail in our public dashboard at \cite{pandametrics}.} 
These dates represent precisely one month before the merge and one month after, which helps us understand the impact of the transition to PoS on validators' performance. 

\begin{figure}
    \centering
    \includegraphics[width=\linewidth]{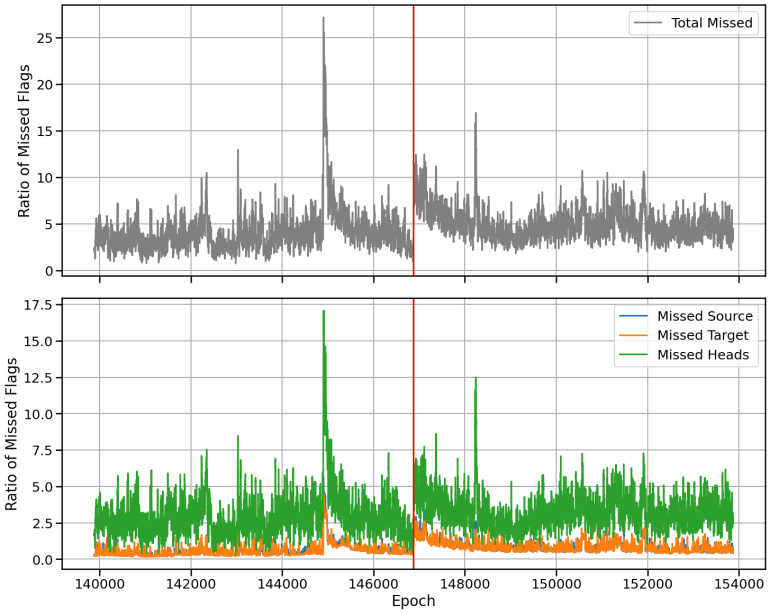}
    \caption{Time distribution of missed flags percentage from the total number of validators. Aggregating the total missed flags on the top chart and distinct missed flags at the bottom.}
    \label{fig:missed-flags}
\end{figure}
\subsection{Descentralization of the Consensus}
\label{subsec:descentralization}
Combining efforts to support the healthy performance of the network is a key point of the Web3 space. We have seen how during PoW, individuals were uniting efforts to be still competitive while supporting decentralization. PoS is not different in that sense. 
We analyzed the level of decentralization pre- and post-merge to determine if the platform improved or got worst in the aspect of decentralization. For that, we looked into the block proposal distribution for the 10 largest entities: mining pools for the pre-merge (determined by fee recipient and using Etherscan~\cite{etherscan}) and staking pools  and entities' validators for the post-merge.
Figure \ref{fig:staking-descentralization} summarizes that distribution, where overall, we observed a rather similar decentralization level. For instance, pre-merge, we see that Ethermine, F2Pool, and Hiveon Pool proposed $28.5\%$, $14.7\%$, and $10.2\%$ of the blocks, respectively. Post-merge, Lido, Coinbase, and Kraken proposed $29.9\%$, $14.2\%$, and $8.4\%$ of the blocks. Nonetheless, it is essential to keep in mind that Lido consists of 28 operators
that are somewhat independent, which adds another level of decentralization. In addition, we observe notably more small-scale/solo staking (``Other” category) participation ($36\%$) in comparison to small-scale/solo mining ($20.9\%$), likely due to the reduced infrastructure and energy requirements that PoS represents over PoW.
 
\subsection{Validators Overall Performance}
\label{subsec:val-performance}
As previously introduced, achieving consensus and the finalization of the chain now depends on the contributions of validators' duties. To rate the performance of validators and ultimately monitor the chain's performance, we looked at the behavior of the overall CL network (CL validators now propose EL blocks). Before making specific comparisons between staking entities, we analyzed the total missed duties of the active validators. 

\subsubsection{Missed Attestation Flags}
\label{subsubsec:missed-flags}
Regarding the validator's attestations, Figure \ref{fig:missed-flags} displays the distribution of missed flags for the $14k$ measured epochs (note the red vertical line that marks the merge epoch). The upper chart in the figure represents the aggregation of all the missed flags normalized by total active validators per epoch. In contrast, the one at the bottom represents the same normalized distribution but is split between each of the flags present in the attestations.     
\begin{figure*}
    \minipage{0.5\textwidth}%
        \includegraphics[width=0.8\linewidth]{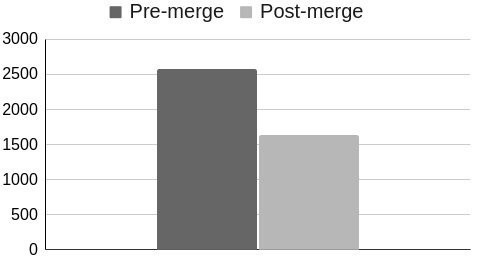}
        \caption{Number of missed blocks pre- and post-merge.}
        \label{fig:missed-blocks}  
    \endminipage\hfill
    \minipage{0.5\textwidth}%
          \includegraphics[width=0.95\linewidth]{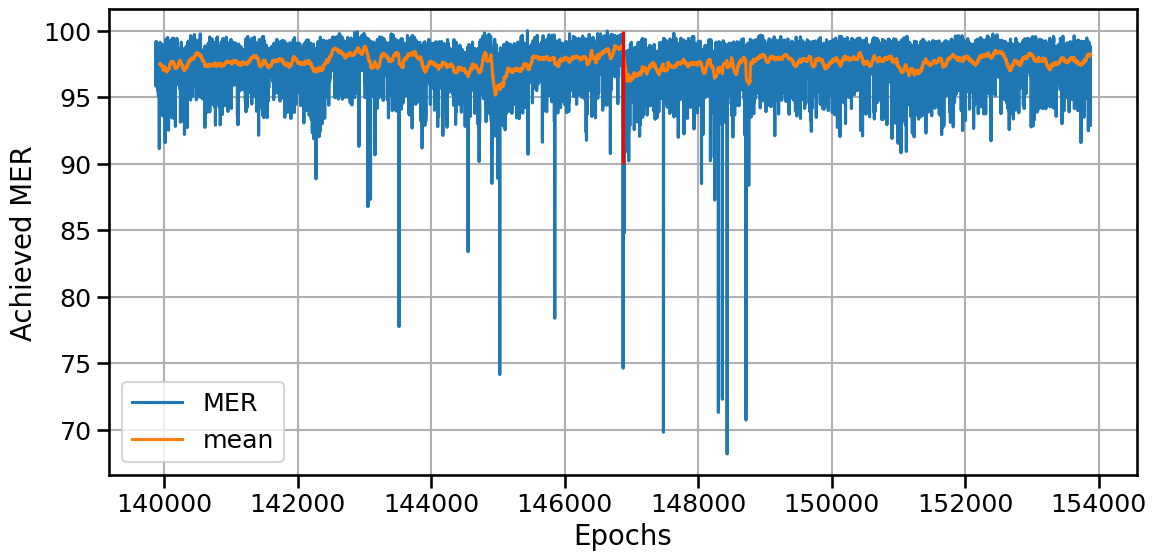}
        \caption{Maximum Extractable Reward for validator network over the $14k$ epochs.}
        \label{fig:mer-ratios}
    \endminipage
\end{figure*}
We observe a slight increase in the total of missed attestation flags, going from a $3.9\%$ pre-merge to a $4.9\%$ post-merge. Considering the different flags included in the attestation, the numbers go from $0.5\%$, $0.5\%$, and $2.9\%$ pre-merge to $0.8\%$, $0.8\%$, and $3.3\%$ post-merge for \emph{Source}, \emph{Target}, and \emph{Head}, respectively.
In the upper graph of Figure \ref{fig:missed-flags}, we can also observe two distinct patterns: 
\begin{enumerate}
    \item A significant spike of $27.17\%$ of missed flags around epoch $145,000$, which we attribute to a large number of restarted clients that were updated to support the merge.
    \item A longer time range of total missed flags between epochs $146,875$ and $148000$ that we attribute to those validators underperforming or missing duties for not having upgraded their software.
\end{enumerate}

Moreover, we can appreciate a large difference in the number of missed flags inside the attestations. In the chart at the bottom of the figure, we can appreciate the validators tend to miss the \emph{Head} flag more often than the \emph{Source} and the \emph{Target} ones. In the chart, we also appreciate that the numbers between the \emph{Source} and the \emph{Target} often match. Meaning that if a peer is missing the \emph{Source} of the attestation (the hash of the justified checkpoint), it is quite likely to miss the \emph{Target} and the \emph{Head} of the attestation. In most cases, missing the three flags in the same attestation can also be interpreted as a ``not attested" sign, which a non-operative client or a non-fully synchronized one might cause. 

\subsubsection{MER}
\label{subsubsec:mer}
The economic incentive promoted by most of the blockchain platforms directly bounds performance and financial rewards. Ethereum is not different in that aspect; validators are rewarded with ETH tokens for their successfully accomplished duties. Figure \ref{fig:mer-ratios} displays the percentage of the MER obtained by the network at each epoch (from attestations and sync committees rewards). We can observe in the figure that the mean for the MER achieved oscillates between $95\%$ and $99\%$, with some sporadic drops that can reach $68\%$ of the MER at some epochs. However, it also means that in each epoch, some validators miss the chance to earn more rewards. As previously introduced in Section \ref{subsubsec:missed-flags}, we attribute these drops to an offline or syncing CL client. 

\subsubsection{Missed Block Proposals}
\label{subsubsec:missed-blocks}
When comparing the number of missed blocks for $7K$ epochs pre-merge to $7K$ epochs post-merge, Figure \ref{fig:missed-blocks} shows an absolute value reduction of $36.40\%$ in missed blocks. We relate this reduction to the fact that missing a block post-merge has a double economic penalty for not getting either CL proposal rewards or EL rewards. Thus, one could expect validators to make every effort to avoid missing block proposals. However, we have to remark that even in the pre-merge scenario, the number of block proposals missed represents $1.13\%$ of the total number of blocks. So the actual $36.40\%$ of reduction refers to a reduction from a $1.13\%$ to a $0.72\%$ ratio of missed blocks.

\subsection{Empirical Distribution of the Rewards in a post-Merge Scenario}
\label{subsec:rewards-empirical-distribution}
From the advantage of indexing all the possible rewards, we have been able to track and compute the different origins for all the validators in the network, and Figure \ref{fig:empirical-rewards-distribution} displays all of them. In the chart, we can appreciate $71.3\%$ of the rewards for a validator are generated through the new consensus layer. Digging a bit more into the CL rewards, the figure showcases that $61.4\%$ of those total rewards are coming attestations, which coincidentally are the most stable rewards a validator can have. Leaving the block proposals in second place with a $7.6\%$ of the total generated rewards, $28.7\%$ for the EL block proposals coming from transaction fees or MEV\footnote{Miner-Extractable Value (MEV) is a new mechanism to compose EL blocks. Block builders compete with each other to build the most rewarding block for the block proposers (maximizing the usage of gas per block, organizing transactions prioritizing tips, or through private transaction channels.)}, and $2.3\%$ for the sync committee rewards.

\subsection{Concatenation of Multiple Block Proposals}
\label{subsec:bp-concatenation}
As long as the beacon chain keeps finalizing, more validators keep joining the network. With a continuous linear increase of $3.8\%$ validators over the two months, the chain went from $428K$ validators at the beginning of the pre-merge period and to $444K$ validators at the end of the post-merge period. 
With this increase in the total number of validators, the chances of being a block proposer decrease. However, the randomness of the RANDAO algorithm still provides some luck to a few validators that can propose multiple blocks in a time range of $2$ months. Table \ref{tab:proposals-per-validator} represents the frequency at which individual validators propose blocks, where we can observe that \updates{$90\%$ of validators proposed either one or no blocks. However, there were still a few lucky proposers that proposed $6$ or $7$ blocks in the same short period. These findings closely align with the fairness rewards study \cite{huang2021rich} performed over different consensus incentive systems, where authors showcase the steady fairness that compounded PoS incentive models, like Ethereum, provide among their participants.}
\begin{table}
    \caption{Table with the number of validators proposing blocks consecutively. (Pre-merge and Post-merge comparison)}
    \label{tab:proposals-per-validator}
    \begin{center}
    \begin{tabular}{ cccccccccc } 
    \hline
         & 0 & 1 & 2 & 3 & 4 & 5 & 6 & 7 \\
    \hline 
        Pre & 255,983 & 131,192 & 34,145 & 6,088 & 805 & 90 & 5 & 1  \\
        Post & 270,251 & 134,458 & 33,682 & 5,747 & 739 & 62 & 5 & -  \\
    \hline
    \end{tabular}
    \end{center}
\end{table}

We have already introduced the origination of staking entities in the space, accumulating thousands of validators. As we already saw, individual validators have chances of proposing several blocks in a short period. And extrapolating those chances to staking entities' validators, there aren't only a higher number of block proposals in short periods but many consecutive block proposals for the entities. Table \ref{tab:consecutive-blocks} shows the number of consecutive blocks proposed for the three biggest staking entities (starting from the 4th consecutive proposal for space reasons). The table shows that Lido, with $27\%$ of the validators, proposed four consecutive blocks $3686$ times in only two months. Achieving even ten consecutive block proposals (a third of the epoch) on five occasions. \updates{Despite each validator eventually gains the ``same" reward as the rest, the uneven token distribution in the network still allows a certain selected group of individuals or entities to accumulate more validators. As works like \cite{kusmierz2022centralized} present, Ethereum's ETH coin shares a Gini coefficient of $0.499$ among the top 100 token holder addresses, outstanding all compared ERC20 token-based blockchains ($0.685$-$0.907$) and remaining close to the BTC Gini coefficient ($0.466$).}

Staking pools or entities have a legit place in the ecosystem; they allow staking for users that would not participate otherwise. However, since the merge happened, having this larger set of chances to propose consecutive blocks could mean a hazard for the overall network. If a large staking entity had some dishonest interests, it could be benefited by allocating users' transactions not only in the order that they want (i.e., performing a sandwich or similar attacks) but to allocate the transactions in the block they want, giving them a higher time window analyze the public transactions' pool. In the ultimate instance, the only thing that prevents large entities from adopting a dishonest position in the network is their public exposure to the same one, as validators are publicly identified in most cases. 

\begin{table}
    \caption{Table with the number of consecutive blocks proposed by the larger staking entities.}
    \label{tab:consecutive-blocks}
    \begin{center}
    \begin{tabular}{ cccccccccc } 
    \hline
        Operator & Merge-Stage & 4 & 5 & 6 & 7 & 8 & 9 & 10 \\
    \hline 
        \multirow{2}{*}{Lido} 
           & Pre & 1849 & 529 & 169 & 51 & 23 & 2 & 5 \\
           & Post & 1837 & 524 & 168 & 48 & 23 & 2 & 4 \\
    \hline
        \multirow{2}{*}{Coinbase} 
           & Pre & 145 & 18 & 4 & - & - & - & - \\
           & Post & 144 & 17 & 4 & - & - & - & - \\
    \hline
        \multirow{2}{*}{Kraken} 
           & Pre & 20 & 3 & 1 & - & - & - & - \\
           & Post & 20 & 3 & 1 & - & - & - & - \\
    \hline
    \end{tabular}
    \end{center}
\end{table}

\subsection{MER per Staking Entities}
\label{subsec:meer-entities}
\begin{figure*}
    \minipage{0.5\textwidth}%
        \includegraphics[width=0.92\linewidth]{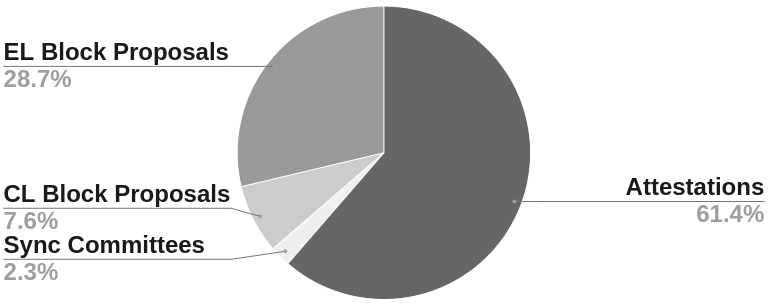}
        \caption{Empirical distribution of rewards from different sources.}
        \label{fig:empirical-rewards-distribution}
    \endminipage\hfill
    \minipage{0.5\textwidth}%
        \includegraphics[width=0.92\linewidth]{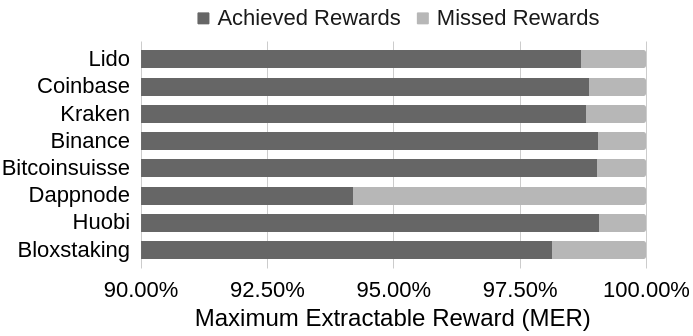}
        \caption{Maximum Extractable Reward for the eight biggest staking entities.}
        \label{fig:mer-per-pool}     
    \endminipage
\end{figure*}

We have already analyzed the entire network's performance in previous sections. However, in a scenario where most users want to delegate their stake to others, Figure \ref{fig:mer-per-pool} displays the achieved MER percentage for the eight biggest pools in the network. 
Considering that MER (max attestations and sync committee rewards) represents $61.4\%$ of the total rewards of the two months we analyzed, we observe that some of the largest staking pools do not fully achieve a portion of the MER. From a minimum of $94,2\%$ for Dappnode to a maximum of $99.07\%$ for Houbi, an average loss of $1.77\%$ of the MER shows a high level of commitment from the respective entities. However, the absolute value of ETH tokens that $1.77\%$ represent for those entities is still relevant.

\section{Conclusion}
\label{sec:conclusion}

This paper presents the shift in the rewards system that ``The Merge" implied for the Ethereum platform. We have unveiled that in the new incentive program that the platform proposes, the consensus layer's rewards represent the 71.3\% of the total rewards, showcasing the importance of having well-behaved and active validators that achieve 61.4\% of their rewards from constant and stable attestations rewards. 

The presented study analyzes the merge of both chains by tracking the missed duties of validators during the process. Overall, the network keeps a healthy behavior that maintains a low ratio of missed blocks and missed attestation flags of 0.72\% and 4.9\%, respectively. We have identified several drops when comparing the obtained CL rewards with the MER during the merge, highlighting some instances where the network miss-performed, achieving 68\% of total possible rewards at some point. 
Moreover, the paper exposes that, currently, the platform faces a high ratio of consecutive blocks being proposed by large staking entities. We introduced that there were even several cases where a third of an epoch was proposed consecutively by a single entity. We have identified the hazard this represents due to these entities' control over which transactions get included and when. 

Ultimately, the paper introduces our rewards indexer tool, which can monitor the obtained rewards over the MER for any set of validators. We have observed that, even though big staking pools show a high commitment to the finalization of the network, they still miss 1,77\% of the rewards generally due to a non-optimal performance on their clients.

As future work to the presented study, we aim to explore the different causes that make validators miss the flags of their attestations and extrapolate the presented rewards study to compare the different existing software.

\section{Acknowledgement}
\label{sec:ack}

This work has been supported by the Lido Ecosystem Grant Organization (LEGO), the Ethereum Foundation under the Research Grant FY21-0356, and Protocol Labs under its Ph.D. Fellowship Program FY22-P2P. We want to thank the blockchain data analyst team from Metrika.co for their insights, feedback, and suggestions on the Execution Layer's rewards analysis. Also, we thank the researchers from Attestant.io for helping with the necessary infrastructure as well as discussions, and in particular, Izzy and Alvaro Revuelta for their constructive feedback on this study.

\bibliographystyle{IEEEtran}
\bibliography{references}

\begin{thebibliography}{10}
\providecommand{\url}[1]{#1}
\csname url@samestyle\endcsname
\providecommand{\newblock}{\relax}
\providecommand{\bibinfo}[2]{#2}
\providecommand{\BIBentrySTDinterwordspacing}{\spaceskip=0pt\relax}
\providecommand{\BIBentryALTinterwordstretchfactor}{4}
\providecommand{\BIBentryALTinterwordspacing}{\spaceskip=\fontdimen2\font plus
\BIBentryALTinterwordstretchfactor\fontdimen3\font minus
  \fontdimen4\font\relax}
\providecommand{\BIBforeignlanguage}[2]{{%
\expandafter\ifx\csname l@#1\endcsname\relax
\typeout{** WARNING: IEEEtran.bst: No hyphenation pattern has been}%
\typeout{** loaded for the language `#1'. Using the pattern for}%
\typeout{** the default language instead.}%
\else
\language=\csname l@#1\endcsname
\fi
#2}}
\providecommand{\BIBdecl}{\relax}
\BIBdecl

\bibitem{eth-whitepaper}
``Ethereum whitepaper,'' \url{ https://ethereum.org/en/whitepaper/}.

\bibitem{cortes2021discovering}
M.~Cortes-Goicoechea and L.~Bautista-Gomez, ``Discovering the ethereum2 p2p
  network,'' in \emph{2021 Third International Conference on Blockchain
  Computing and Applications (BCCA)}.\hskip 1em plus 0.5em minus 0.4em\relax
  IEEE, 2021, pp. 81--88.

\bibitem{hildenbrandt2018kevm}
E.~Hildenbrandt, M.~Saxena, N.~Rodrigues, X.~Zhu, P.~Daian, D.~Guth, B.~Moore,
  D.~Park, Y.~Zhang, A.~Stefanescu \emph{et~al.}, ``Kevm: A complete formal
  semantics of the ethereum virtual machine,'' in \emph{2018 IEEE 31st Computer
  Security Foundations Symposium (CSF)}.\hskip 1em plus 0.5em minus 0.4em\relax
  IEEE, 2018, pp. 204--217.

\bibitem{wohrer2018smart}
M.~Wohrer and U.~Zdun, ``Smart contracts: security patterns in the ethereum
  ecosystem and solidity,'' in \emph{2018 International Workshop on Blockchain
  Oriented Software Engineering (IWBOSE)}.\hskip 1em plus 0.5em minus
  0.4em\relax IEEE, 2018, pp. 2--8.

\bibitem{oliveira2022analysis}
P.~H.~F. Oliveira, D.~M. Rezende, H.~S. Bernardino, S.~M. Villela, and A.~B.
  Vieira, ``Analysis of account behaviors in ethereum during an economic impact
  event,'' \emph{arXiv preprint arXiv:2206.11846}, 2022.

\bibitem{roughgarden2020transaction}
T.~Roughgarden, ``Transaction fee mechanism design for the ethereum blockchain:
  An economic analysis of eip-1559,'' \emph{arXiv preprint arXiv:2012.00854},
  2020.

\bibitem{saleh2021blockchain}
F.~Saleh, ``Blockchain without waste: Proof-of-stake,'' \emph{The Review of
  financial studies}, vol.~34, no.~3, pp. 1156--1190, 2021.

\bibitem{randao-code}
``Ethereum's rng randao,'' \url{https://github.com/randao/randao}.

\bibitem{paris-hardfork}
``Ethereum's paris hard fork,'' \url{https://ethereum.org/en/history/#paris}.

\bibitem{salimitari2017profit}
M.~Salimitari, M.~Chatterjee, M.~Yuksel, and E.~Pasiliao, ``Profit maximization
  for bitcoin pool mining: A prospect theoretic approach,'' in \emph{2017 IEEE
  3rd international conference on collaboration and internet computing
  (CIC)}.\hskip 1em plus 0.5em minus 0.4em\relax IEEE, 2017, pp. 267--274.

\bibitem{nakamoto2008bitcoin}
S.~Nakamoto, ``Bitcoin whitepaper,'' \emph{URL: https://bitcoin. org/bitcoin.
  pdf-(: 17.07. 2019)}, 2008.

\bibitem{simon2021block}
J.~R. Simon and K.~Geetha, ``Block mining reward prediction with polynomial
  regression, long short-term memory, and prophet api for ethereum blockchain
  miners,'' in \emph{ITM Web of Conferences}, vol.~37.\hskip 1em plus 0.5em
  minus 0.4em\relax EDP Sciences, 2021, p. 01004.

\bibitem{albert2020gasol}
E.~Albert, J.~Correas, P.~Gordillo, G.~Rom{\'a}n-D{\'\i}ez, and A.~Rubio,
  ``Gasol: gas analysis and optimization for ethereum smart contracts,'' in
  \emph{International Conference on Tools and Algorithms for the Construction
  and Analysis of Systems}.\hskip 1em plus 0.5em minus 0.4em\relax Springer,
  2020, pp. 118--125.

\bibitem{daian2019flash}
P.~Daian, S.~Goldfeder, T.~Kell, Y.~Li, X.~Zhao, I.~Bentov, L.~Breidenbach, and
  A.~Juels, ``Flash boys 2.0: Frontrunning, transaction reordering, and
  consensus instability in decentralized exchanges,'' \emph{arXiv preprint
  arXiv:1904.05234}, 2019.

\bibitem{qin2022quantifying}
K.~Qin, L.~Zhou, and A.~Gervais, ``Quantifying blockchain extractable value:
  How dark is the forest?'' in \emph{2022 IEEE Symposium on Security and
  Privacy (SP)}.\hskip 1em plus 0.5em minus 0.4em\relax IEEE, 2022, pp.
  198--214.

\bibitem{cortes2021resource}
M.~Cortes-Goicoechea, L.~Franceschini, and L.~Bautista-Gomez, ``Resource
  analysis of ethereum 2.0 clients,'' in \emph{2021 3rd Conference on
  Blockchain Research \& Applications for Innovative Networks and Services
  (BRAINS)}.\hskip 1em plus 0.5em minus 0.4em\relax IEEE, 2021, pp. 1--8.

\bibitem{liu2019survey}
Z.~Liu, N.~C. Luong, W.~Wang, D.~Niyato, P.~Wang, Y.-C. Liang, and D.~I. Kim,
  ``A survey on applications of game theory in blockchain,'' \emph{arXiv
  preprint arXiv:1902.10865}, 2019.

\bibitem{london-hardfork}
``Ethereum's london hard fork,'' \url{https://ethereum.org/en/history/#london}.

\bibitem{altair-hardfork}
``Ethereum's altair hard fork,'' \url{https://ethereum.org/en/history/#altair}.

\bibitem{eth-cl-specs}
``Ethereum's consensus specs,''
  \url{https://github.com/ethereum/consensus-specs}.

\bibitem{buterin2017casper}
V.~Buterin and V.~Griffith, ``Casper the friendly finality gadget,''
  \emph{arXiv preprint arXiv:1710.09437}, 2017.

\bibitem{state-analyzer}
M.~Cortes-Goicoechea, ``Ethereum consensus layer's state analyzer,''
  \url{https://github.com/cortze/eth-cl-state-analyzer}, 2022.

\bibitem{pandametrics}
M.~Labs, ``Pandametrics - ethereum rewards' performance visualizer,''
  \url{https://pandametrics.xyz}, 2022.

\bibitem{etherscan}
``Etherscan, ethereum blockchain explorer,'' \url{https://etherscan.io/}.

\bibitem{huang2021rich}
Y.~Huang, J.~Tang, Q.~Cong, A.~Lim, and J.~Xu, ``Do the rich get richer?
  fairness analysis for blockchain incentives,'' in \emph{Proceedings of the
  2021 International Conference on Management of Data}, 2021, pp. 790--803.

\bibitem{kusmierz2022centralized}
B.~Kusmierz and R.~Overko, ``How centralized is decentralized? comparison of
  wealth distribution in coins and tokens,'' in \emph{2022 IEEE International
  Conference on Omni-layer Intelligent Systems (COINS)}.\hskip 1em plus 0.5em
  minus 0.4em\relax IEEE, 2022, pp. 1--6.

\end{thebibliography}

\end{document}